\documentclass[preprint,showpacs,preprintnumbers,amsmath,amssymb,superscriptaddress]{revtex4}

\usepackage{graphicx}
\usepackage{dcolumn}
\usepackage{bm}
\usepackage{color}
\usepackage{longtable}
\newcommand{\lb}{\left(}
\newcommand{\rb}{\right)}
\newcommand{\ls}{\left[}
\newcommand{\rs}{\right]}

\newcommand{\br}{\mbox{\boldmath$r$}}
\newcommand{\hatr}{\hat{\br}}

\newcommand{\bJ}{\mbox{\boldmath$J$}}

\newcommand{\alp}{\alpha}

\newcommand{\del}{\delta}
\newcommand{\eps}{\epsilon}
\newcommand{\vep}{\varepsilon}

\newcommand{\lam}{\lambda}

\newcommand{\sig}{\sigma}

\newcommand{\vph}{\varphi}

\newcommand{\Gam}{\Gamma}
\newcommand{\Del}{\Delta}

\newcommand{\half}{\frac{1}{2}}

\newcommand{\benu}{\begin{enumerate}}
\newcommand{\eenu}{\end{enumerate}}
\newcommand{\beq}{\begin{equation}}
\newcommand{\eeq}{\end{equation}}
\newcommand{\beqn}{\begin{eqnarray}}
\newcommand{\eeqn}{\end{eqnarray}}

\newcommand{\beqd}{\begin{eqnarray*}}
\newcommand{\eeqd}{\end{eqnarray*}}
\newcommand{\bea}{\begin{array}}
\newcommand{\eea}{\end{array}}
\newcommand{\bcen}{\begin{center}}
\newcommand{\ecen}{\end{center}}
\newcommand{\btab}{\begin{tabular}}
\newcommand{\etab}{\end{tabular}}
\newcommand{\bsub}{\begin{subequations}}
\newcommand{\esub}{\end{subequations}}

\newcommand{\beit}{\begin{itemize}}
\newcommand{\enit}{\end{itemize}}

\newcommand{\rto}{\rightarrow}

\def\m@thcombine#1#2{%
  \setbox0=\hbox{$#1$}
  \setbox1=\hbox{$#2$}
  \ifdim\wd0>\wd1
    \setbox0=\hbox to\wd1{\hss\box0\hss}
  \else
    \setbox1=\hbox to\wd0{\hss\box1\hss}
  \fi
  \mathop{\vcenter{
    \offinterlineskip\box0\box1}}}
\def\lesim{\m@thcombine<\sim}
\def\gesim{\m@thcombine>\sim}

\begin{document}

\title{
Persistent contribution of unbound
quasiparticles to the pair correlation in continuum
Skyrme-Hartree-Fock-Bogoliubov approach}

\author{Y. Zhang}
\affiliation{State Key Laboratory of Nuclear Physics and Technology, School of Physics, Peking University, Beijing 100871, China}
\affiliation{Graduate School of Science and Technology, Niigata University, Niigata 950-2181, Japan}
\author{M. Matsuo}
\affiliation{Graduate School of Science and Technology, Niigata University, Niigata 950-2181, Japan}
\affiliation{Department of Physics, Faculty of Science, Niigata University, Niigata 950-2181, Japan}
\author{J. Meng}\thanks{e-mail: mengj@pku.edu.cn}
\affiliation{School of Physics and Nuclear Energy Engineering, Beihang University, Beijing 100191, China}
\affiliation{State Key Laboratory of Nuclear Physics and Technology, School of Physics, Peking University, Beijing 100871, China}
\affiliation{Department of Physics, University of Stellenbosch, Stellenbosch, South Africa}

\begin{abstract}
The neutron pair correlation in nuclei near the neutron drip-line is
investigated using the selfconsistent continuum
Skyrme-Hartree-Fock-Bogoliubov theory formulated with the
coordinate-space Green's function technique. Numerical analysis is
performed for even-even $N=86$ isotones in the Mo-Sn region, where
the $3p_{3/2}$ and $3p_{1/2}$ orbits lying near the Fermi energy are
either weakly bound or unbound. The quasiparticle states originating
from the $l=1$ orbits form resonances with large widths, which are
due to the low barrier height and the strong continuum coupling
caused by the pair potential. Analyzing in detail the pairing
properties and roles of the quasiparticle resonances, we found that
the $l=1$ broad quasiparticle resonances persist to feel the pair
potential and contribute to the pair correlation even when their
widths are comparable with the resonance energy.

\end{abstract}


\maketitle

\section{\label{sec:intr}Introduction}
Many interests have been taken recently on pairing properties of
neutron-rich nuclei near the drip-line. The most peculiar case could
be the firstly observed halo nucleus, $^{11}$Li, where two neutrons
forming the halo would not be bound to the nucleus if the pair
correlation were absent~\cite{Tanihata,Hansen,Esbensen,Meng-Li}.
Similar examples are also suggested in nuclei near the neutron
drip-line in heavier mass region: e.g., possible giant halo
structure (consisting of more than two neutrons) which is predicted
in Ca and Zr isotopes
 by the selfconsistent mean-field models~\cite{Meng-Zr,Meng-Ca,GrassoGhalo}.
A new aspect in these examples is that the pair correlation occurs among
neutrons occupying unbound or weakly bound orbits
 with low angular momentum $l=0$ or $1$ whose wave functions extend
far outside the nucleus due to the low (zero) centrifugal barrier.

A useful theoretical tool to study the pair correlation in the
weakly bound nuclei in all the mass regions, except the lightest
ones, is the coordinate-space Hartree-Fock-Bogoliubov (HFB)
approach~\cite{DobHFB1,Belyaev,Bulgac,MengRCHB1998}, in which the
quasiparticle wave functions of weakly bound and unbound nucleons
are described in the coordinate-space representation. Indeed the
pairing properties in nuclei near the drip-line have been studied
extensively within the HFB
scheme~\cite{GrassoGhalo,DobHFB2,Grasso,DD-Dob,MMS05,Yamagami05,
Hamamoto03,Hamamoto04,Hamamoto05,Hamamoto06,Oba} as well as the
relativistic Hartree-(Fock)-Bogoliubov
models~\cite{Meng-Li,Meng-Zr,Meng-Ca,MengRCHB1998,Meng2006,Zhou2003,Zhou2010,Long2010}.
The pair correlation we have to deal with here is, however, a rather
complex and unresolved problem, and there exist controversial issues
concerning the roles of weakly bound and unbound orbits.  For
instance, it has been argued in
Refs.~\cite{Hamamoto03,Hamamoto04,Hamamoto05,Hamamoto06} that
neutrons in the weakly bound and unbound orbits with $l=0$ and $1$
tend to decouple from the pair field generated by the other neutrons
because of the large spatial extension of their wave functions. It
is also claimed that those neutrons contribute very little to the
total pair correlation in nuclei. On the contrary, other studies
show large pairing effects even on the weakly bound neutrons,
leading to the pairing anti-halo effect~\cite{Bennaceur} and the
increase of the neutron pairing gap for weaker binding of neutrons
or for shallower neutron Fermi energy~\cite{Yamagami05,Oba}.

In this paper,
we would like to present an investigation of the pairing
properties of nuclei close to the neutron drip-line, with intentions to clarify
the roles of weakly bound and unbound orbits with low angular momenta.

To perform
this study, there exist some physically and technically important points which need to be treated carefully.
Firstly, precise description of the wave functions
outside the
nucleus must be guaranteed since we
deal with weakly-bound and unbound orbits. We achieve it in the present study by
using the coordinate-space mesh representation for the Skyrme-Hartree-Fock-Bogoliubov
model~\cite{DobHFB1}.

Secondly, also related to the first point, a suitable boundary condition
needs to be imposed  on the wave functions of the quasiparticle states in the continuum,
which also have sizable contribution to the pair correlation in the case of nuclei
with a shallow Fermi energy close to zero. Note here that
the quasiparticle states whose excitation energy exceed the separation energy form
the continuum spectrum because they couple to
scattering waves~\cite{DobHFB1,Bulgac,Belyaev}.
The Hartree-Fock single-particle orbits emerge
as resonant quasiparticle states with finite width~\cite{Belyaev}.
To describe this situation, one needs to guarantee the
asymptotic form of the quasiparticle wave function as the scattering wave.
In this way, we can avoid artificial discretization
of the continuum spectrum, and can evaluate the width of a resonant
quasiparticle state. This allows us to investigate how
the resonant quasiparticle states contribute to
 the pair correlation.

Thirdly, it is important to describe selfconsistently the pair potential,
which is the key quantity describing the pair correlation. To achieve this, however,
the continuum quasiparticle states including both resonant
and non-resonant states are to be summed up in evaluating the one-body densities.
We adopt the Green's function technique~\cite{Belyaev} that provides a simple and effective way of summing.
Thus, we are able to perform in the present work the fully selfconsistent
continuum Hartree-Fock-Bogoliubov calculations, i.e., we derive
selfconsistently both the Hartree-Fock potential (using the Skyrme functional) and the pair potential
(using a density-dependent contact interaction as the effective pairing force).
It  is noted that
the theoretical framework of the present analysis shares many common aspects with
that in Refs.~\cite{Hamamoto03,Hamamoto04,Hamamoto05,Hamamoto06}, but we differ
in that we utilize the self-consistent pair potential as well as the Hartree-Fock mean-field.
Our approach is rather similar to the continuum Skyrme-HFB calculations formulated
in Refs.~\cite{Grasso,GrassoGhalo,Fayans}.

We will perform numerical analysis for
the $N=86$ isotones in the Mo-Sn region. The Skyrme-HFB theory with the
parameter set SLy4 suggests the presence of weakly-bound neutron single-particle orbits
 above the $N=82$ shell gap in neutron rich nuclei with $N\gesim 84$ and $Z\lesim 50$.
In the $N=86$ isotones, particularly,
 the $3p_{1/2}$ and $3p_{3/2}$ orbits emerge close to the zero energy with
the Hartree-Fock single-particle energies ranging from
 $\vep \sim -0.5 $~MeV to unbound
resonances around $\vep \sim 0.7$~MeV.
 Hence it provides us with a good testing ground
to study the role of weakly bound low-$l$ orbits in the pair correlation.

In the numerical analysis, we shall pay special attentions to  the pair
density (called also the pairing tensor or the abnormal density in the literature)
and the related quantities.
The pair density is one of the most relevant quantities to the pair correlation since
it shows up in the definitions of both the selfconsistent pair potential and
the pair correlation energy. Using this quantity we will show that
the $l=1$ weakly bound or unbound
orbits keep finite and sizable contribution to the pair correlation
even when they form very broad quasiparticle resonances, and when they are
located above the potential barrier. The pairing gap associated with
the unbound quasiparticle
states are also found to stay finite. This result is different from
those in
Refs.\cite{Hamamoto03,Hamamoto04,Hamamoto05,Hamamoto06}, which suggest
the decoupling of the $l=0, 1$ weakly bound and unbound orbits from the pair correlation.
We shall discuss the origin of the difference.

Finally, we remark that
the present analysis is related
very closely to Ref.~\cite{Oba}, where, however, the Hartree-Fock potential
is replaced by a simple Woods-Saxon potential although the deformation effect is
taken into account. In the present work, we do not discuss the deformation effect for
simplicity, but instead we investigate in detail the roles of weakly bound
and unbound orbits especially with the low angular momentum
by using the fully selfconsistent continuum
Hartree-Fock-Bogoliubov
calculations assuming the spherical symmetry.

In section II, we describe the formulation of
 the continuum Skyrme-HFB theory using the Green's function
technique. After presenting the results including the numerical
details and the related discussions in section III, we draw
conclusions in Section IV.

%
\section{\label{sec:form}Formalism}
%
 \subsection{\label{subsec:HFB}Hartree-Fock-Bogoliubov equation with Skyrme interaction}

  In the Hartree-Fock-Bogoliubov (HFB) theory, the pair correlated nuclear system is
  described in terms of the independent quasiparticles.  The HFB equation for
  the quasiparticle wave function $\phi_i(\br\sig)$ in the coordinate space is
  \beq
      \int d\br'\sum_{\sig'}
      \left(
        \begin{array}{cc}
          h(\br\sig,\br'\sig')-\lam\del(\br-\br')\del_{\sig\sig'}
        & \tilde{h}(\br\sig,\br'\sig') \\
          \tilde{h}^*(\br\tilde{\sig},\br'\tilde{\sig}')
        & -h^*(\br\tilde{\sig},\br'\tilde{\sig}')+\lam\del(\br-\br')\del_{\sig\sig'} \\
        \end{array}
      \right)\phi_i(\br'\sig') = E_i \phi_i(\br\sig)
  \eeq
  where $E_i$ is the quasiparticle energy, and $\lam$ is the
  chemical potential or the Fermi energy.  The Hartree-Fock Hamiltonian
  $h$ and the pair Hamiltonian $\tilde{h}$ can be obtained by
  the variation of the total energy functional with respect to
  the particle density matrix $\rho(\br\sig, \br'\sig')$ and
  pair density matrix $\tilde{\rho}(\br\sig, \br'\sig')$ respectively.
  The two density matrices can be combined in a generalized density matrix $\mathcal{R}$ as
  \beqn
    \mathcal{R}(\br\sig,\br'\sig') &\equiv &
     \left(
       \begin{array}{cc}
         \rho(\br\sig,\br'\sig') & \tilde{\rho}(\br\sig,\br'\sig') \\
         \tilde{\rho}^*(\br\tilde{\sig},\br'\tilde{\sig}')
         & \del(\br-\br')\del_{\sig\sig'}- \rho^*(\br\tilde{\sig},\br'\tilde{\sig}')\\
       \end{array}
     \right), \label{eq:genR-def-1}
  \eeqn
  where the particle density matrix $\rho(\br\sig,\br'\sig')$
  and pair density matrix $\tilde{\rho}(\br\sig,\br'\sig')$ are just the "11"
  and "12" components of $\mathcal{R}$ respectively.

  The energy density functional of the Skyrme interaction is
  constructed with the local densities, such as the particle density $\rho(\br)$,
  the kinetic-energy density $\tau(\br)$, and the spin-orbit density $\bJ(\br)$, etc.,
  defined with the particle density matrix $\rho(\br\sig, \br'\sig')$~\cite{ENGEL1975NPA,Bender2003}.
  We adopt a density dependent delta interaction (DDDI) for the $p$-$p$ channel:
 \beq
   v_{\mbox{\tiny pair}}(\br,\br') = \half (1-P_{\sig}) V_0\ls 1-\eta\lb
   \frac{\rho(\br)}{\rho_0}\rb^{\alp}\rs \del(\br-\br'),
   \label{eq:DDDI-V}
 \eeq
 which presents similar properties as the pairing interaction with
 finite range~\cite{Meng1998PRC-pair}.
 Then the pair Hamiltonian $\tilde{h}$ is reduced to the local pair potential
 \beq
   \Del(\br) = \half V_0\ls 1-\eta\lb\frac{\rho(\br)}{\rho_0}\rb^{\alp}\rs \tilde{\rho}(\br),
   \label{eq:DDDI}
 \eeq
 where the local pair density $\tilde{\rho}(\br)$ is defined with the pair density matrix
 $\tilde{\rho}(\br\sig,\br\sig)$~\cite{DobHFB1}.

  For the spherical symmetry, the generalized density matrix $\mathcal{R}$
  can be expanded on the spinor spherical harmonics as
  \beq
    \mathcal{R}(\br\sig,\br'\sig') = \sum_{ljm} Y_{ljm}(\hatr\sig)\mathcal{R}_{lj}(r,r')Y^*_{ljm}(\hatr'\sig').
    \label{eq:genR-expand}
  \eeq
  Using the 11 and 12 components of $\mathcal{R}_{lj}$,
  one can write the radial local densities as
\begin{subequations}
 \beqn
   \rho(r) &=& \frac{1}{4\pi}\sum_{lj} (2j+1)\mathcal{R}^{11}_{lj}(r,r), \label{eq:radial-rho}\\
   \tau(r) &=&  \frac{1}{4\pi}\sum_{lj}(2j+1)
        \ls \overrightarrow{\frac{d}{dr}} \mathcal{R}^{11}_{lj}(r,r')\overleftarrow{\frac{d}{dr'}} +
        l(l+1) \frac{\mathcal{R}^{11}_{lj}(r,r')}{rr'} \rs_{r=r'}, \label{eq:radial-tau} \\
  J(r) &=&  \frac{1}{4\pi r}\sum_{lj}(2j+1)\ls j(j+1) -l(l+1) -\frac{3}{4}\rs \mathcal{R}^{11}_{lj}(r,r),
       \label{eq:radial-J}\\
  \tilde{\rho}(r) &=& \frac{1}{4\pi}\sum_{lj} (2j+1)\mathcal{R}^{12}_{lj}(r,r), \label{eq:radial-rhot}
 \eeqn \label{eq:density-radial}
\end{subequations}
 where $\overleftarrow{\displaystyle\frac{d}{dr'}}$ denotes the derivative operator with
 respect to $r'$ acting from right to left.

  The quasiparticle wave function is represented as
  \beq
    \phi_i(\br\sig) = \frac{1}{r} \phi_{lj}(r) Y_{ljm}(\hatr\sig),~~
    \mbox{where}~~ \phi_{lj}(r) = \left(
                                    \begin{array}{c}
                                     \vph_{1,lj}(r) \\
                                     \vph_{2,lj}(r)
                                    \end{array}
                                  \right),
  \eeq
 which obeys the radial HFB equation
  \beq
    \left(
      \begin{array}{cc}
        -\displaystyle\frac{d}{dr}\frac{\hbar^2}{2m^*}\frac{d}{dr} + U_{lj}(r) - \lam
      & \Del(r) \\
        \Del(r)
      & \displaystyle\frac{d}{dr}\frac{\hbar^2}{2m^*}\frac{d}{dr} - U_{lj}(r) + \lam  \\
      \end{array}
    \right) \phi_{lj}(r,E) = E \phi_{lj}(r,E). \label{eq:HFB-Eq-radial}
  \eeq
The explicit expressions of the effective mass $m^*_q$ and
the Hartree-Fock potential $U_{lj}(r)$ can be found in Ref.
\cite{DobHFB1,ENGEL1975NPA}, and they are constructed by
the radial local densities (\ref{eq:density-radial}) and their derivatives.

 \subsection{\label{subsec:SHFBGF}HFB Green's function and densities with correct asymptotic behavior}

 In the conventional Skyrme HFB theory, one solves the radial HFB equation (\ref{eq:HFB-Eq-radial})
 with the box boundary condition
 $\phi_{lj}(r,E)=0$ at the edge  of the box  $r=R$ ($R$ being the box size)
to obtain the discretized eigen solutions
 for the single-quasiparticle energy and the corresponding wave functions.
 Then the generalized density matrix $\mathcal{R}$ can be constructed
 by a sum over discretized quasiparticle states.
 Although the box boundary condition is appropriate for the deeply bound states, it is not
 suitable for the weakly bound and the continuum states unless a large enough box size is taken.

 Here the Green's function method is used to impose the correct asymptotic behaviors
 on the wave functions especially for
 the continuum states, and to calculate the densities.

The HFB Green's function $\mathcal{G}_{0,lj}(r,r',E)$
 can be constructed with  solutions of the radial HFB equation (\ref{eq:HFB-Eq-radial}).
 Suppose $\phi^{(rs)}_{lj}(r,E)$ and
 $\phi_{lj}^{(+s)}(r,E)~(s=1,2)$ are independent solutions
 of the HFB equation (\ref{eq:HFB-Eq-radial})
 that satisfy the boundary conditions at
 the origin and at the edge  of the box, $r=R$, respectively,
 then the HFB Green's function is given
 by~\cite{Belyaev,Matsuo01}
  \beq
    \mathcal{G}_{0,lj}(r,r',E) = \sum_{s,s'=1,2} c^{ss'}_{lj}(E)
    \ls \theta(r-r') \phi_{lj}^{(+s)}(r,E)\phi_{lj}^{(rs')T}(r',E) +
     \theta(r'-r) \phi_{lj}^{(rs')}(r,E)\phi_{lj}^{(+s)T}(r',E)\rs.
    \label{eq:HFBGF-def-5}
  \eeq
The coefficients $c^{ss'}_{lj}(E)$ are expressed in terms of the Wronskians as
  \beq
    \left(
      \begin{array}{cc}
        c^{11}_{lj} & c^{12}_{lj} \\
        c^{21}_{lj} & c^{22}_{lj} \\
      \end{array}
    \right) = \left(
                \begin{array}{cc}
                  w_{lj}(r1,+1) & w_{lj}(r1,+2) \\
                  w_{lj}(r2,+1) & w_{lj}(r2,+2) \\
                \end{array}
              \right)^{-1} \label{HFB-wronskian-coeff}
  \eeq
  with
  \beqn
    w_{lj}(rs,+s') &=& \frac{\hbar^2}{2m}
    \ls \vph^{(rs)}_{1,lj}(r)\frac{d}{dr}\vph^{(+s')}_{1,lj}(r)
    - \vph^{(+s')}_{1,lj}(r)\frac{d}{dr}\vph^{(rs)}_{1,lj}(r) \right. \nonumber \\
    && \left. -\vph^{(rs)}_{2,lj}(r)\frac{d}{dr}\vph^{(+s')}_{2,lj}(r)
    + \vph^{(+s')}_{2,lj}(r)\frac{d}{dr}\vph^{(rs)}_{2,lj}(r) \rs.  \nonumber \\
    \label{HFB-wronskian}
  \eeqn

  To impose the correct asymptotic behavior on the wave function
  for the continuum states, we adopt the boundary condition
  as follows,
  \beq
    \left\{
    \begin{array}{ll}
      \phi^{(rs)}_{lj}(r,E): & \mbox{regular at the origin $r=0$} \\
      \phi_{lj}^{(+s)}(r,E): & \mbox{outgoing wave at $r\rto \infty$}
    \end{array}
    \right. \label{eq:conitnuum-bc}
  \eeq
Explicitly, the solutions $\phi_{lj}^{(+s)}(r,E)$ at $r > R$ satisfy
 \beq
    \phi_{lj}^{(+1)}(r,E) \rto \left(
                                 \begin{array}{c}
                                   e^{ik_+(E)r} \\
                                   0 \\
                                 \end{array}
                               \right), ~~~~
    \phi_{lj}^{(+2)}(r,E) \rto \left(
                                 \begin{array}{c}
                                    0 \\
                                   e^{ik_-(E)r}\\
                                 \end{array}
                               \right).  \label{eq:bbc}
  \eeq
  Here $k_{\pm}(E) = \sqrt{2m(\lam\pm E)}/\hbar$ and their branch cuts are
  chosen so that $\text{Im} k_{\pm}>0$ is satisfied.

\begin{figure}[!h]
    \centering
    \includegraphics[scale=0.5]{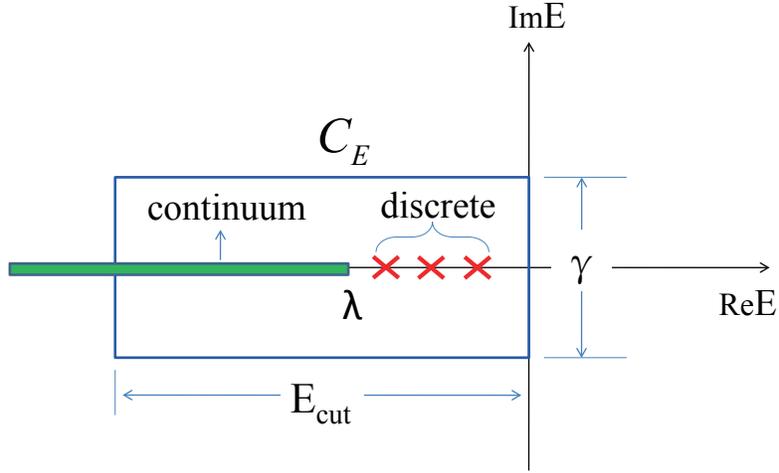}
    \caption{
    (color online) Contour path $C_E$ to perform the integration of the
    Green's function on the complex quasiparticle energy plane.  The path is
    chosen to be a rectangle with the height $\gamma$ and the length
    $E_{\text{cut}}$.  The crosses denote the discrete quasiparticle states.
    The continuum states are denoted by the solid stripe below the Fermi energy $\lam$.}
    \label{fig:contourchk}
  \end{figure}

   The generalized density matrix can be
 obtained by the contour integral of the Green's function
 on the complex quasiparticle energy plane, which in the
 spherical case can be written as~\cite{Belyaev,Matsuo01}
  \beq
    \mathcal{R}_{lj}(r,r')=  \frac{1}{2\pi i}\oint_{C_{E}} dE~ \frac{\mathcal{G}_{0,lj}(r,r',E)}{rr'}.
    \label{HFBden-GF-radial}
  \eeq
 The contour $C_E$ should be chosen to enclose the negative
 energy part of the quasiparticle spectra,
 as shown in
  Figure~\ref{fig:contourchk}, so that all the quasiparticle states inside the
  contour are summed up.
  Here the discrete quasiparticle
  states are denoted by crosses above the Fermi energy $\lam$.
  Below the Fermi energy, the continuum quasiparticle states
  are denoted by the solid stripe.
 As a result, the radial local densities (\ref{eq:density-radial}) can be
 calculated by the contour integral of the radial Green's function.
 In this way, we realize a fully selfconsistent continuum
 Skyrme HFB calculations.

%
\subsection{\label{sec:details}Numerical details}
%

 For the Skyrme interaction, we choose the parameter set SLy4~\cite{Sly41998},
 and for the pairing interaction the DDDI parameters in Eq.~(\ref{eq:DDDI}) are adopted as
 $V_0=-458.4$~MeV~fm$^{-3}$, $\eta=0.71$, $\alp=0.59$,
 and $\rho_0=0.08$~fm$^{-3}$~\cite{Matsuo2006PRC,Matsuo2007NPAc,Matsuo2010PRC}.
 They reproduce the experimental neutron pairing gap along the Sn isotopic chain.
 We remark also that the parameter $V_0$ is chosen in such a way that the DDDI
 reproduces the $^{1}S$ scattering length $a=-18.5$ fm of the bare nuclear force
 in the low density limit $\rho(r) \rightarrow 0$, i.e., in the free space outside
 the nucleus. Because of this constraint and the density-dependence,
 the pairing interaction strength is
 large around the surface, and even increases in the exterior.
 The truncation of the quasiparticle states is up to the
 angular momentum $l=12$ and $j=25/2$ and to the maximal quasiparticle
 energy $E_{\text{cut}}=60$~MeV.

  For the contour integration of the Green's function,
  the path $C_E$ is chosen to be a rectangle as shown in
  Figure~\ref{fig:contourchk} with
  the height $\gamma=1$~MeV and the length $E_{\text{cut}}=60$~MeV,
  which symmetrically encloses the real negative quasiparticle energy axis.
  For the contour integration we adopt
  an energy step $\Delta E=0.01$~MeV on the contour path.
  We have checked that for the choice of these contour path parameters,
  the precision for $\rho(r)$ and $\tilde{\rho}(r)$ thus obtained are up to
  $10^{-10}~\text{fm}^{-3}$ and $10^{-9}~\text{fm}^{-3}$ respectively.
  We choose the box size $R=20$~fm, and
  the mesh size $\Delta r=0.2$~fm.  We have also checked that dependence of the
  results on the box size is very small thanks to
  the boundary condition (\ref{eq:bbc}) with proper asymptotic form.

%
\section{\label{sec:resu}Results and discussion}
%

 In this section, taking the isotonic chain $N=86$ as an example,
 we will discuss in detail how characters of
the weakly bound and unbound quasiparticle states of neutrons
vary as the neutron Fermi energy approaches zero, and how they
contribute to the pair correlation.

 \subsection{HFB ground states and the quasiparticle spectra}

Some properties of the HFB ground state  for $^{136}$Sn, $^{134}$Cd,
$^{132}$Pd, $^{130}$Ru, and $^{128}$Mo are listed in the first four rows
in Table~\ref{tab:spectrum-N86}. The neutron Fermi energy (the first row)
monotonically increases from $-2.39$ MeV in $^{136}$Sn to $-0.42$ MeV in
 $^{128}$Mo as the proton number $Z$ decreases. This is because the neutron
Hartree-Fock potential becomes shallower as $Z$ decreases, and we could not find a
self-bound HFB solution (with $\lambda <0$ ) in $^{126}$Zr and lighter isotones.

Table~\ref{tab:spectrum-N86} also shows the total neutron pair correlation energy
  \beqn
    E_{\text{pair}} &=& \half \int d\br~\Del(\br)\tilde{\rho}(\br),
    \label{eq:Epair-1}
  \eeqn
and the average pairing gap
 \beqn
   \Del_{uv} &=& \frac{\int d\br~\Del(\br)~\tilde{\rho}(\br)}
   {\int d\br~\tilde{\rho}(\br)}. \label{eq:avgap-uv}
 \eeqn
The quantity in the denominator is the total neutron 'pair number'
 \beq
  \tilde{N} = \int \tilde{\rho}(\br)d\br \label{eq:Ntilde-def},
 \eeq
which represents the
amount of the pair condensate.
It is noted that the variation of
the average pairing gap and the total pair correlation energy
along the isotones is less than $10\%$ from $^{136}$Sn to the last bound nucleus $^{128}$Mo.

It is useful to investigate properties of individual quasiparticle states
which are the elementary mode of single-particle motion in the HFB theory
and  the building blocks of the pair density. It is noted that the spectrum of the quasiparticles, i.e., the eigenstates of the
HFB equation, includes  both the discrete
 ($0<E<|\lam|$) and continuum ($E>|\lam|$) quasiparticle states.
 Accordingly, the pair density can be expressed as a sum of
 contributions from individual quasiparticle states as
  \beq
   \tilde{\rho}(r) = \sum_{nlj,E_{nlj}<|\lam|} \tilde{\rho}_{nlj}(r) +
   \sum_{lj}\int^{E_{\text{cut}}}_{|\lam|}\tilde{\rho}_{lj}(r,E)~dE,
 \eeq
where the first term in r.h.s. is the sum over the discrete states, and the second term
represents the integral of
the contribution to the pair density $\tilde{\rho}(r)$ from the continuum
quasiparticle state with quantum number $lj$ at energy $E$  .
If we include the discrete quasiparticle states in the definition of
$\tilde{\rho}_{lj}(r,E)$,
the above equation can be expressed as
 \beq
   \tilde{\rho}(r) = \sum_{lj} \tilde{\rho}_{lj}(r),
   \text{~where~} \tilde{\rho}_{lj}(r)=\int^{E_{\text{cut}}}_{0}\tilde{\rho}_{lj}(r,E)~dE,
   \label{eq:rhotr-rhotE}
 \eeq
and $\tilde{\rho}_{lj}(r,E)$ can
be calculated as
 \beq
   \tilde{\rho}_{lj}(r,E) =\left(\frac{2j+1}{4\pi r^2}\right)\frac{1}{\pi}\text{Im}
   \mathcal{G}_{0,lj}^{(12)}(r,r,-E-i\eps). \label{eq:rhotildeE-def}
 \eeq

 We can also calculate  contributions from the state with quantum number $lj$ at energy $E$
 to the pair number $\tilde{N}$ as
 \beq
  \tilde{n}_{lj}(E) = \int 4\pi r^2\tilde{\rho}_{lj}(r,E)dr, \label{eq:ntildeE-def}
 \eeq
 which satisfies
 \beq
   \tilde{N} = \sum_{lj} \int^{E_{\text{cut}}}_0 \tilde{n}_{lj}(E) dE.
 \eeq
 We call the quantity $\tilde{n}_{lj}(E)$ the 'pair number density'
in the following.  We can also
 calculate the 'occupation number density' ${n}_{lj}(E) = \int 4\pi r^2{\rho}_{lj}(r,E)dr$,
 which is discussed in Refs.~\cite{Hamamoto03,Hamamoto04,Hamamoto05,Hamamoto06,Grasso,Fayans}.
 In the following,
 we will investigate the pair number density
 $\tilde{n}_{lj}(E)$ rather than the occupation number density $n_{lj}(E)$ since
 we found that the pair number density $\tilde{n}_{lj}(E)$ represents more clearly
 the structure of continuum quasiparticle states relevant to the pair correlation.

 With the smoothing parameter $\eps$ in Eq.~(\ref{eq:rhotildeE-def})
  the $\del$ function (no width) originating from a
 discrete quasiparticle state is simulated by a Lorentzian function
 with the full width at half maximum (FWHM) of $2\eps$. In the following calculation,
 we take $\eps=5$~keV to discuss the structure of pair number density.
 Subtracting the smoothing width $2\eps=10$ keV from the FWHM of the peak,
 we obtain the physical width $\Gamma$ of the peak.

\begin{figure}[!h]
\includegraphics[width=8cm]{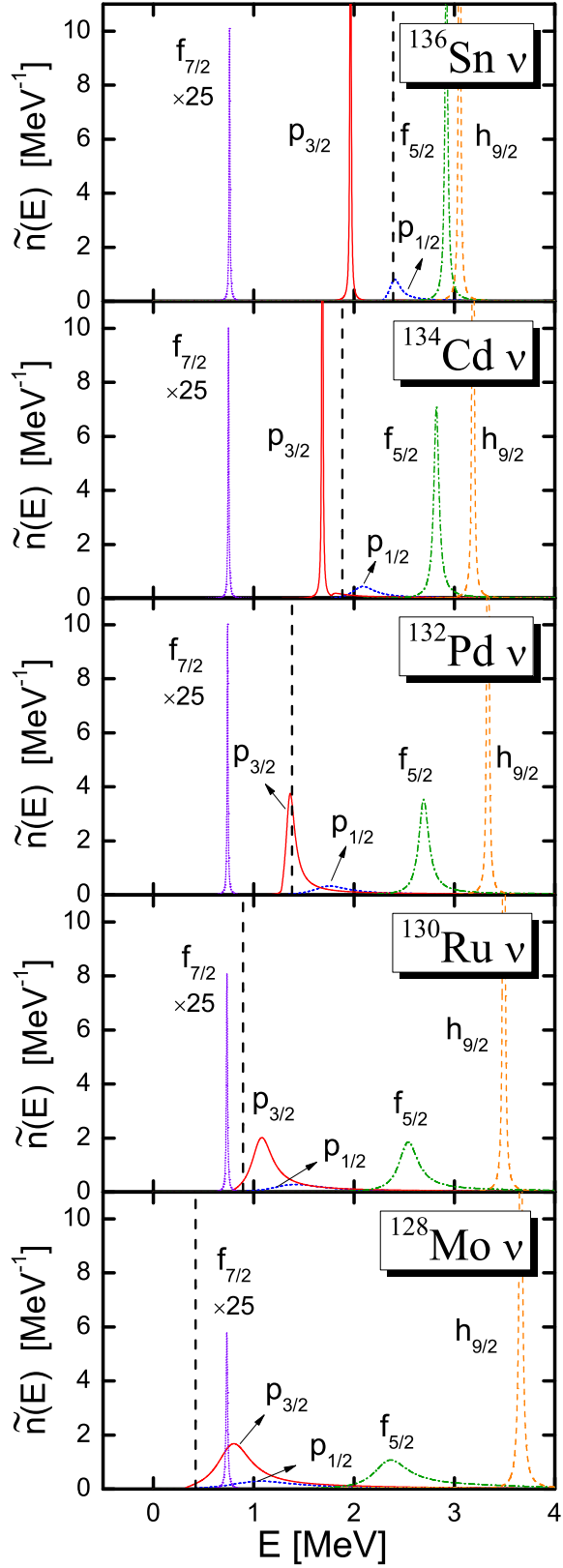}\\
\vspace{-0.5cm}
\caption{Pair number densities $\tilde{n}_{lj}(E)$ of
 neutron quasiparticle states with different $lj$ around the threshold energy in the $N=86$ isotones, obtained
 with the self-consistent continuum Skyrme HFB theory using Green's function method.
 The thick dashed line denotes the threshold energy $|\lam|$ for the
 continuum quasiparticle states. The density of state for $f_{7/2}$
 is divided by a factor of $25$.}\label{fig:qpspectrumN86II}
\end{figure}

 Figure~\ref{fig:qpspectrumN86II} shows the pair number densities $\tilde{n}_{lj}(E)$
 for neutron quasiparticle states in a low-lying energy interval $E=0\sim 4$ MeV in the $N=86$ isotones.
 A peak structure below the threshold energy  $E=|\lam|$ (the dashed vertical line) is
 a discrete quasiparticle
 state (simulated by the Lorentzian function), and a peak
 above the dashed line may be identified as a quasiparticle resonance.
 The width $\Gamma$ of the peak as well as the peak energy $E_{\text{q.p.}}$ are
tabulated in
 Table~\ref{tab:spectrum-N86}.
 In the same table, we also show the corresponding Hartree-Fock single-particle
 energy $\vep$,
 which is the eigen solution of the Hartree-Fock single-particle Hamiltonian $h$
 obtained with the box boundary condition.
 An eigen state with $\vep<0$ is discrete (bound)
 single-particle orbit, and $\vep>0$ is discretized continuum single-particle orbit
 whose energy gives an estimate for the Hartree-Fock
 single-particle resonance energy.

 In $^{136}$Sn, the quasiparticle states corresponding to weakly-bound Hartree-Fock single-particle
 states, $2f_{7/2}$ and $3p_{3/2}$,  are discrete states
 with no width (less than $0.1$~keV in the actual numerical calculation) as they are
 located below the threshold energy $|\lam|$.
 The Hartree-Fock single-particle state $3p_{1/2}$ is already in the continuum
($\vep \sim 70$ keV), and
 forms a quasiparticle resonance located just above
 the threshold energy with $E_{\text{q.p.}}\sim |\lam| + 15$ keV with a large width $\Gam=122$~keV. On the contrary,
 the peaks for higher angular momentum states, e.g., $2f_{5/2}$ and $1h_{9/2}$, have finite but
 smaller widths, forming narrow quasiparticle resonances.

 As the proton number decreases, the neutron potential becomes shallower,
 and both the Fermi energy and
 the single-particle energies are raised up.
 The width grows larger in $^{134}$Cd than in $^{136}$Sn for all the quasiparticle resonances above the threshold, although the weakly bound single-particle
 states $2f_{7/2}$ and $3p_{3/2}$ remain the discrete states in the
 quasiparticle spectrum.

 When the Fermi energy is raised up further in $^{132}$Pd,
 the  $3p_{3/2}$ quasiparticle state becomes a resonance located
 just around the
 threshold energy, with a large width $95$~keV. Since
 the corresponding Hartree-Fock state is still bound (although
 the binding energy is very small, $\vep \sim - 10$ keV), it is the pair correlation
 that makes the corresponding quasiparticle state unbound with
 the finite width.

In $^{130}$Ru,  the single-particle energies
are raised further. All the single-particle orbits under discussion
except $2f_{7/2}$ become unbound. Correspondingly the widths of the quasiparticle
resonances $3p_{3/2}$, $3p_{1/2}$, $2f_{5/2}$ and $1h_{9/2}$ grow also.

 When we come to the last bound nucleus $^{128}$Mo in this isotonic chain,
 all the quasiparticle states lie in the continuum.
 It is noticed that both the $3p_{3/2}$ and $3p_{1/2}$ quasiparticle states
 become very broad resonances whose width is comparable
 with the resonance energy $E_{\text{q.p.}}-|\lam|$ measured from
 the threshold.
 The quasiparticle state $2f_{7/2}$ is now a quasiparticle resonance
 embedded in the continuum although the Hartree-Fock single-particle state $2f_{7/2}$ is still a bound orbit.
 The width
 comes from the continuum coupling caused by the pair correlation, as
  the $3p_{3/2}$ state in $^{132}$Pd. The continuum coupling effect is
 much larger for the $3p_{3/2}$ state than $2f_{7/2}$, as is seen in the
 considerably different values of the width.

 \subsection{ Pairing effects on the resonance width}

 The large widths of the $3p_{1/2}$ and $3p_{3/2}$ quasiparticle resonances have two origins. One is
 the barrier penetration of the Hartree-Fock plus centrifugal potential, which is
 low for the states with low angular momentum (the barrier height of the $p_{1/2}$ and $p_{3/2}$ states
 in $^{128}$Mo is {$0.351$} and {$0.344$} MeV, respectively). This is effective even
 without the pair correlation. The other is the presence of
 the pair potential, because of which even a bound single-particle orbit can couple with
 continuum states, then forms a quasiparticle resonance.
 The latter effect may be seen by examining the
 pair number density $\tilde{n}_{lj}(E)$ under a variation of the
 strength of the effective pairing interaction.

 Table~\ref{tab:spectrum-128Mo-fp}
 shows the quasiparticle
 resonance widths in $^{128}$Mo obtained with different
 pairing interaction parameter $\eta = 0.84, 0.71$ and $0.62$ in Eq.~(\ref{eq:DDDI-V}).
 It is seen that for the changes of the average pairing gap
$\Delta_{uv} = 0.41 \sim 0.68 \sim 1.08$ MeV, the variation of the widths of the
  $3p_{1/2}$ and $3p_{3/2}$ quasiparticle resonances are $\Gam=0.73 \sim 0.80 \sim 0.94$~MeV
  and $0.34 \sim 0.40 \sim 0.49$~MeV, respectively.
  We deduce that
 the pairing effect on the widths are approximately $\sim 100$ keV for $3p_{1/2}$,
 and a slightly smaller for $3p_{3/2}$.
 Comparably large pairing effect is also seen in the width of
 $2f_{5/2}$ quasiparticle resonance where the corresponding Hartree-Fock single-particle state
 is a broad resonance already without the pair correlation.
 (On the contrary, the pairing effect is not large for
 the other quasiparticle resonances
 arising from the hole orbits and those with narrow resonances.)
 The pair correlation increases significantly the
 width of the quasiparticle resonances corresponding to weakly bound orbits with low angular momentum
 or to the Hartree-Fock single-particle resonances close to the barrier
 top. Since the wave functions of these quasiparticle resonances
 have significant amplitude in the barrier region,
 the influence of the pair potential on the continuum coupling can be
 effective.

\subsection{Contribution of continuum quasiparticle states to the pair correlation}

 Let us now investigate how the quasiparticle resonances,
shown in Fig.~\ref{fig:qpspectrumN86II}, contribute to the neutron pair
correlation.

\begin{figure}[!h]
\includegraphics[width=8cm]{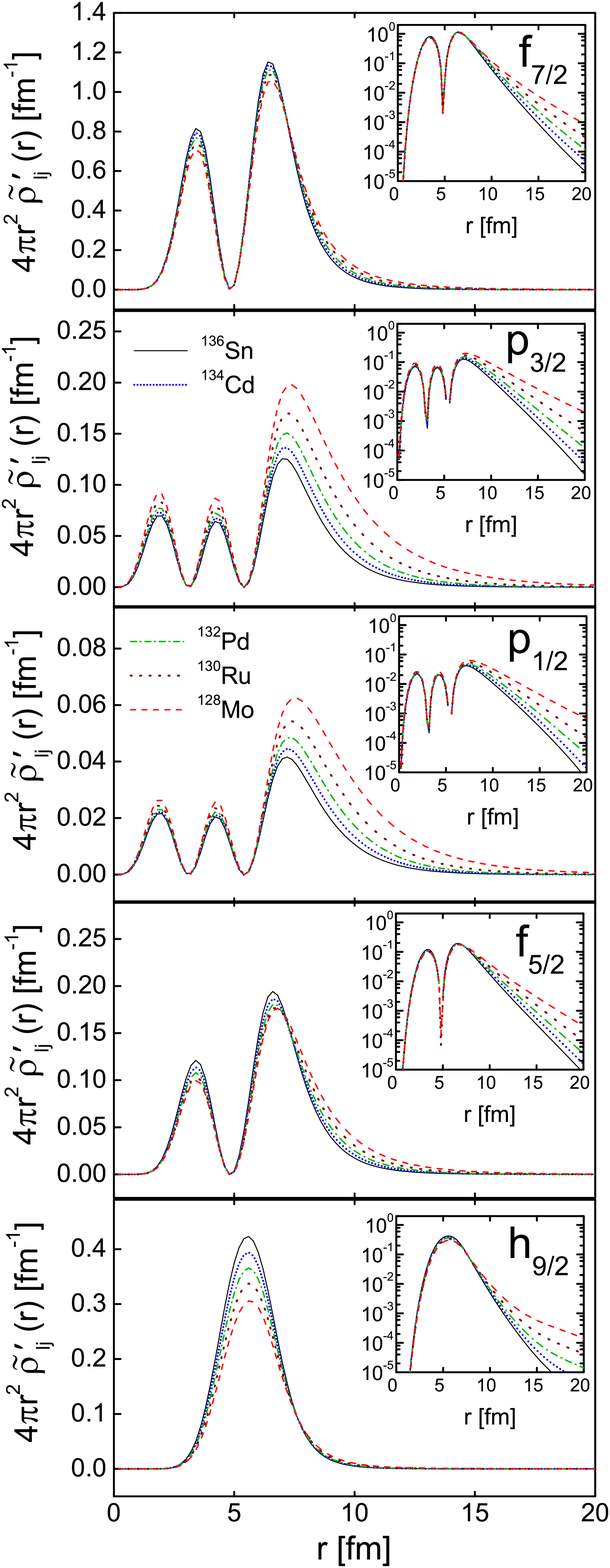}\\
\caption{Neutron pair density $4\pi r^2\tilde{\rho}_{nlj}'(r)$
 contributed by the low-lying quasiparticle states shown in Fig.~\ref{fig:qpspectrumN86II}
 for the $N=86$ isotones,
 where $\tilde{\rho}_{nlj}'(r)=\int^{4~\text{MeV}}_0 dE~\tilde{\rho}_{lj}(r,E)$.
 The inserts present the same density distribution in a log scale. } \label{fig:rhotNjlII}
\end{figure}

 For this purpose, we examine their contribution to the
 neutron pair density in the low-lying energy interval $E=0 \sim 4$ MeV.
 We denote $\tilde{\rho}_{nlj}'(r)$ for the partial contribution from the low-lying quasiparticle state, and evaluate it by performing the integral in Eq.~(\ref{eq:rhotr-rhotE}) with $E_{\text{cut}}=4$~MeV
 for the quasiparticle resonances
$2f_{7/2}, 3p_{3/2}, 3p_{1/2}$ etc. The quantity weighted with
the volume element, $4\pi r^2\tilde{\rho}_{nlj}'(r)$, is shown
in Figure~\ref{fig:rhotNjlII} for the $N=86$ isotones.

As the nucleus becomes more and more weakly bound from $^{136}$Sn to $^{128}$Mo,
 a significant variation of the pair density $\tilde{\rho}_{nlj}'(r)$ is seen for the
 $3p_{3/2}$ and $3p_{1/2}$ states:  the amplitude of $\tilde{\rho}_{nlj}'(r)$
increases dramatically at the positions far outside the surface, $r \approx 7 - 15 $~fm.
For the $3p_{1/2}$ state, the increase at $r=8$ and $10$~fm are $80\%$ and $200\%$, respectively
while the increase inside, e.g. at $r=2$ fm, is only $\sim 20\%$.
The other quasiparticle states $2f_{7/2}, 2f_{5/2}$ and $1h_{9/2}$ exhibit a similar trend
of extending outside but to a much weaker extent.

Evaluating the volume integral of $\tilde{\rho}_{nlj}'(r)$, we list in Table~\ref{tab:spectrum-N86} the quantity
  $\tilde{N}_{nlj}' = \int 4\pi r^2\tilde{\rho}_{nlj}'(r)~dr $, which represents a contribution
to the pair number $\tilde{N}$ from the low-lying quasiparticle states.
 It is clear that the pair number  $\tilde{N}_{nlj}'$ of the $3p_{3/2}$ and $3p_{1/2}$ states
in the most weakly bound $^{128}$Mo is twice as big as those in $^{136}$Sn, as a result of the obvious increase of $\tilde{\rho}_{nlj}'(r)$ at the positions $r\approx 7-15$ fm.

\begin{figure}[!h]
\includegraphics[width=8cm]{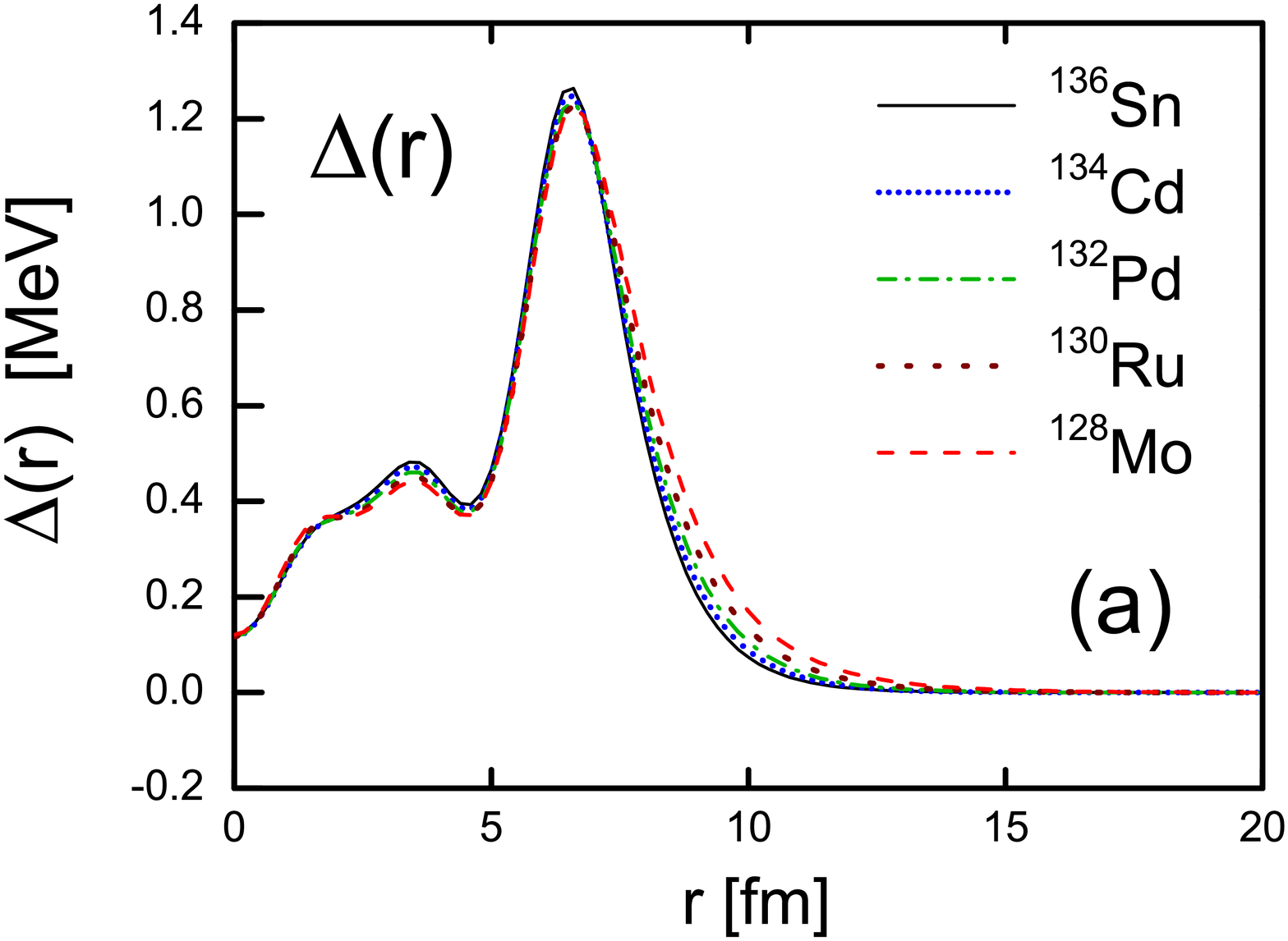}\\
\includegraphics[width=8cm]{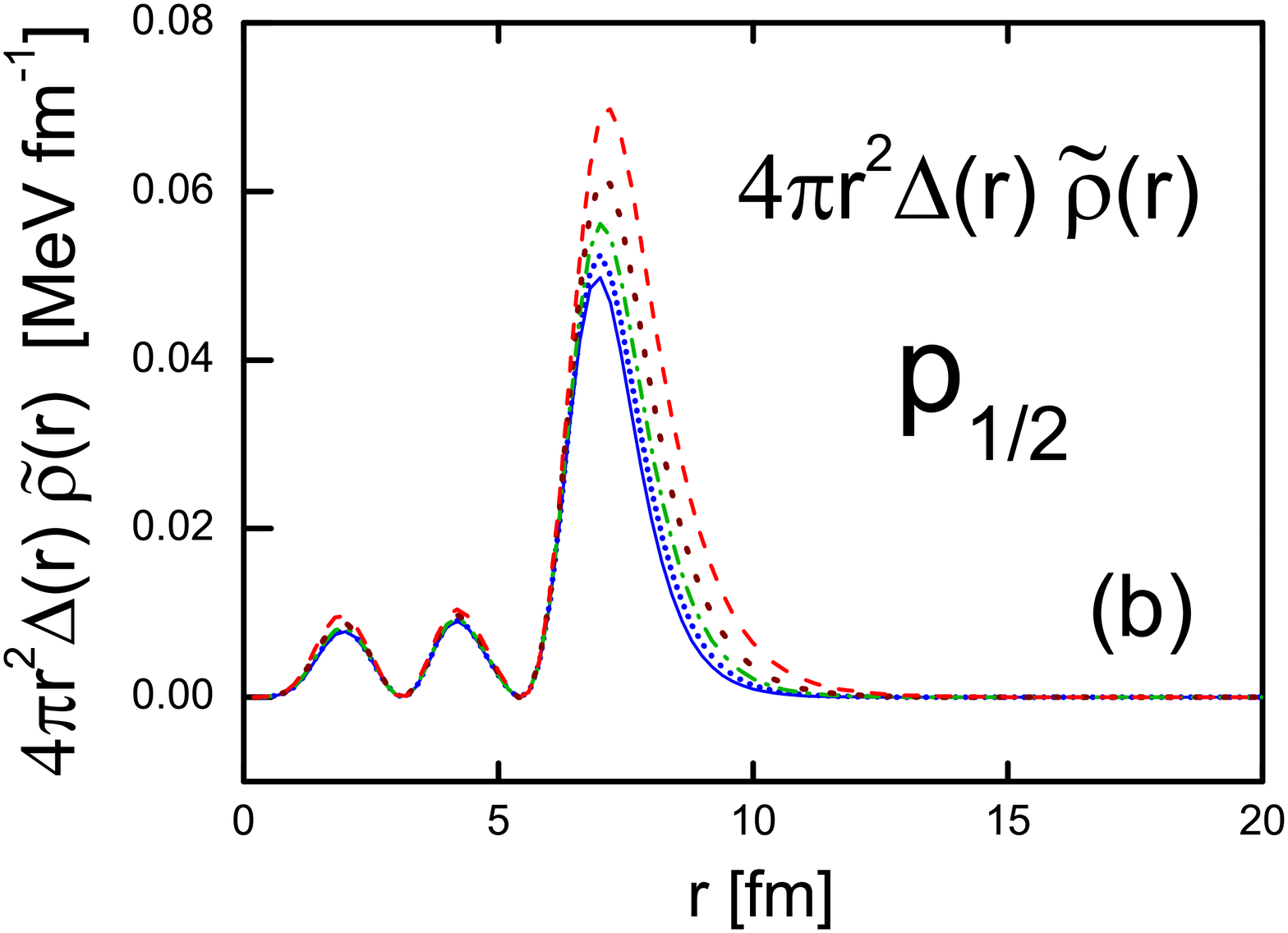}\\
\caption{(a) Neutron pair potential $\Del(r)$ in the $N=86$ isotones.
 (b) Integrand of the pair correlation energy, $4\pi r^2\tilde{\rho}_{nlj}'(r) \Del(r)$,
 for the $3p_{1/2}$ quasiparticle state in the $N=86$ isotones,
 where the pair density $\tilde{\rho}_{nlj}'(r)$ is the
 contribution from the $3p_{1/2}$ resonant quasiparticle state which is shown in
 Figure~\ref{fig:rhotNjlII}.} \label{fig:dpotrhot-p12}
\end{figure}

We show also in Table~\ref{tab:spectrum-N86} a partial contribution
 \beqn
    E_{\text{pair},nlj}' &=& \half \int 4\pi r^2 dr~\Del(r)\tilde{\rho}_{nlj}'(r)
    \label{eq:partial-pairenergy}
 \eeqn
 to the pair correlation energy from the low-lying quasiparticle state.
 It can be seen that, moving from $^{136}$Sn to $^{128}$Mo,
 the contributions of the $3p_{1/2}$ and $3p_{3/2}$ states do increase up to around 70\%. It is clear  that
this increase of the pair correlation energy is due to not only the increase in $\tilde{\rho}_{nlj}'(r)$
but also the large spatial overlap between the pair density $\tilde{\rho}_{nlj}'(r)$
and the pair potential $\Delta(r)$. In Figure~\ref{fig:dpotrhot-p12} we show
the neutron pair potential $\Del(r)$ and the product
of the pair potential and the pair density $4\pi r^2 \Del(r) \tilde{\rho}_{nlj}'(r) $,
the integrand of the
 pair correlation energy, Eq.~(\ref{eq:partial-pairenergy}).
Since the radial profile of the pair potential $\Delta(r)$ is not only surface-peaked
but also extends up to $r\sim 10$ fm, the two quantities have
significant overlap, and hence  the product $4\pi r^2 \Del(r) \tilde{\rho}_{nlj}'(r) $
exhibits a significant increase at $r\approx 7-10$ fm. This brings about large
increase of the pair correlation energy.

 We note here that the energies of the Hartree-Fock single-particle orbits
 corresponding to the  $3p_{3/2}$ and $3p_{1/2}$ states move upward around and beyond the threshold:
  from $\vep=-0.48$~MeV (weakly
 bound) to $\approx 0.36$~MeV (unbound with the single-particle energy comparable with the
barrier height $\approx 0.34$ MeV) in the $3p_{3/2}$ case,
 and from $\approx 0.07$~MeV (around the threshold) to $\approx 0.68$~MeV
 (even above the barrier height $\approx 0.35$ MeV) in the $3p_{1/2}$ case.
 In the least bound case ($^{128}$Mo), the  $3p_{3/2}$ and $3p_{1/2}$ quasiparticle resonances
 have the widths $\Gamma=0.396$ and $0.802$ MeV, respectively, which are
  comparable with the resonance energy. Nevertheless,
 both the pair number $\tilde{N}_{nlj}'$ and the pair correlation energy $E_{\text{pair},lj}'$ continue to
 increase as is seen above. This indicates that the weakly bound and unbound states can feel
  the pair potential and contribute to the pair correlation in sizable way.

\subsection{Effective pairing gap of continuum quasiparticle states}

 In order to make quantitative estimate for the influence of
 the pair potential on the low-lying quasiparticle states,
 we  evaluate state dependent effective pairing gap which can be defined by
  \beq
    {\Del}_{uv, nlj}' = \frac{\int 4\pi r^2 \Del(r)\tilde{\rho}_{nlj}'(r)~dr}
   {\int 4\pi r^2 \tilde{\rho}_{nlj}'(r)~dr} =
   - 2E_{\text{pair},nlj}'/\tilde{N}_{nlj}'
   \label{eq:effgapuv-def-1}
 \eeq
 using the pair density $\tilde{\rho}_{nlj}'(r)$ for the
 specific quasiparticle state. We list it
  in Table~\ref{tab:spectrum-N86}.

 In the nucleus $^{136}$Sn, the effective pairing gaps of the $3p_{3/2}$ and
$3p_{1/2}$ states are
 ${\Del}_{uv, nlj}'=0.591$~MeV and $0.570$~MeV, respectively, which are about $77 \sim 80\%$ of
the total-average pairing gap $\Del_{uv}^{\text{tot.}}=0.736$~MeV.
The effective pairing gaps
 still keep finite values of the same order
in the last bound nucleus $^{128}$Mo, where
 the $3p_{1/2}$ and $3p_{3/2}$ Hartree-Fock orbits are both unbound.  For the $3p_{3/2}$ state,
the effective pairing gap is
 ${\Del}_{uv, nlj}'=0.529$~MeV. It  stays at
 the level of $78\%$ of the average gap
$\Del_{uv}^{\text{tot.}}=0.678~\text{MeV}$.
 Even for the $3p_{1/2}$ resonance with large width, the effective pairing gap
 $0.501$~MeV keeps $74\%$ of the total.
 The variation of the effective pairing gaps of the $3p$ states from $^{136}$Sn
 to $^{128}$Mo is also small, i.e., they decrease only slightly by $\sim 10\%$.

 The facts that the effective pairing gaps of the   $3p_{3/2}$ and
$3p_{1/2}$ states are slightly smaller than the total average value, and
that they decrease as the orbits become less bound and become unbound in the continuum,
can be ascribed to the decoupling effect~\cite{Hamamoto03,Hamamoto04,Hamamoto05,Hamamoto06},
 which is expected to originate from the possible small overlap between the
 single-particle wave function and the pair potential.
 In Ref.~\cite{Hamamoto05}, the effective pairing
 gap in the weakly bound $p$ orbit
 is suggested to be less than $50\%$ of the average, and
 possibly less than $1/3$ for an unbound $p$ orbit. Compared with these numbers,
 the decoupling effect observed here ($\approx 20-25\%$) is much smaller.
 Namely, we can see that
 these quasiparticle states persist to feel the pair potential
 and contribute to the pair correlation even if they become unbound and have
 large width.

 The difference between the conclusions of
 our analysis and those of Ref.~\cite{Hamamoto05}
 can be explained as follows.  In our analysis,
 the self-consistent pair potential not only peaks around
 the surface ($r \approx 5-7$ fm), but also extends outside (up to $10$~fm or more)
 as shown in Figure~\ref{fig:dpotrhot-p12}~(a),
 whereas the pair potential in Ref.~\cite{Hamamoto03,Hamamoto04,Hamamoto05,Hamamoto06} has a Woods-Saxon shape
 whose main part is concentrated inside the nucleus.
 Meanwhile, the pair density $4\pi r^2\tilde{\rho}_{nlj}'(r)$
 also peaks around the surface and extends outsides.
 Let us take, for instance,
 the $3p_{1/2}$ resonance state in $^{128}$Mo whose Hartree-Fock single-particle energy is around $0.7$~MeV.
 Considering the wave function of the quasiparticle state at the peak energy $E_{\text{q.p.}}$,
 its upper component $\varphi_1(r,E_{\text{q.p.}})$ has large amplitude around the barrier
 as the state is located above the barrier height~($0.35$~MeV),
 and it oscillates in the asymptotic region.
  On the other hand, the lower
 component $\varphi_2(r,E_{\text{q.p.}})$ exhibits an exponentially decaying asymptotics
 $\propto \exp(-\kappa r)$ with
 $\kappa = \sqrt{2m (|\lambda| + E_{\text{q.p.}})}/\hbar$~\cite{DobHFB2}.
Since the contribution of this state to the pair density
 $\tilde{\rho}_{nlj}'(r)$ is given by the product of $\varphi_1(r,E_{\text{q.p.}})$  and
 $\varphi_2(r,E_{\text{q.p.}})$,  the pair density  is confined around the nucleus
 ($r \lesssim 15$~fm in the present numerical examples) and the largest
 amplitude of $4\pi r^2\tilde{\rho}_{nlj}'(r)$ shows up around the surface even though
 the quasiparticle state is located far above the threshold and has a large width.
 Consequently, such states have sizable overlap with the pair potential
 and thus keep finite effective pairing gap even when
 the nucleus becomes more weakly bound.

 Conversely we may argue a condition for occurrence of the strong decoupling
 in a semi-quantitative way. Since the spatial extension of the pair density
 is characterized by a size constant
$r_{\tilde{\rho}}\equiv 1/\kappa = \hbar/\sqrt{2m (|\lambda| + E_{\text{q.p.}})}$,
 the strong decoupling can be expected when
 $r_{\tilde{\rho}} \gg R_{\text{surf.}}$, i.e. only
 when both the Fermi energy $|\lam|$ and the quasiparticle energy $E_{\text{q.p.}}$ are sufficiently small.
 Here  the energy $E_{\text{q.p.}}$ of a discrete or
 resonance quasiparticle state has a lower bound $E_{\text{q.p.}} \gesim \Delta_{nlj}'$ given by the
 effective pairing gap.
 For the $p$ states in $^{128}$Mo, we find $E_{\text{q.p.}}\sim 0.7$~MeV, and $|\lambda| > 0.4$~MeV,
 therefore, $r_{\tilde{\rho}}\sim 4$~fm, which is not much larger than  the nuclear
radius $R_{\text{surf.}}\sim 5$~fm.
 This explains why the decoupling is weak here.


\section{\label{summary}Conclusions}


We have investigated the neutron pair correlation in
neutron-rich nuclei with small neutron separation energies
by means of the fully selfconsistent continuum Skyrme HFB theory,
in which the Green's function method is utilized to describe precisely
the asymptotic behavior of scattering waves for the unbound quasiparticle
states in the continuum.
We have clarified
how weakly-bound and unbound neutron orbits contribute to the
pair correlation properties, especially the orbits with the
low angular momentum $l=1$ which have large spatial extensions.
We have chosen the even-even $N=86$ isotones in the
Sn-Mo region for numerical analysis, and investigated
in detail the pairing properties associated with the neutron
$3p_{3/2}$ and $3p_{1/2}$ orbits, whose Hartree-Fock single-particle energies (resonances)
vary in the interval of
$-0.5$ MeV $< \vep < 0.7$ MeV, covering both weakly-bound
and unbound cases.

We found the following features from the numerical analysis.
When the $3p$ quasiparticle states are embedded in the
continuum above the threshold, they immediately become broad resonances with
large widths. This is because the barrier height of the Hartree-Fock plus centrifugal
potential is low for the $p$ orbits
($\sim 0.35$ MeV in the present examples), and also because the pair potential
which remains effective around the barrier region gives rise to additional
coupling to the scattering wave in the exterior. The numerical results show that the width of
the quasiparticle resonances of the $3p$ states are comparable to the excitation
energy measured from the threshold. In spite of such large width ($\Gam \sim 1$ MeV),
  the contribution of the broad quasiparticle resonances
to the pair correlation remains finite or can even increase. We found that
the effective pairing gaps of the
broad quasiparticle resonances have a comparable size to the total average
pairing gap, indicating that the continuum quasiparticle states persist to contribute to the
pair correlation. To be more precise, there exist some reduction of the effective
pairing gap of $20-25$\% from the total average gap. However,
this reduction of the effective pair gap is much smaller than what is discussed
in Ref.~\cite{Hamamoto05}.

Summarizing, even the broad quasiparticle $p$-wave resonances in the continuum do contribute to the
pair correlation as far as it is located not far from the Fermi energy. This is different
from the decoupling scenario~\cite{Hamamoto03,Hamamoto04,Hamamoto05,Hamamoto06}. The reason
for the difference is that the pair correlation in the present study is described
selfconsistently using the effective pairing interaction which has enhancement outside
the nuclear surface, and in this case the pair potential is enhanced largely around
the surface and proximate exterior,
keeping overlap with the low-$l$ resonant quasiparticle states.

\begin{acknowledgments}

The work was partly supported by the Grant-in-Aid for Scientific
Research (Nos. 20540259, 21105507 and 21340073) from the Japan
Society for the Promotion of Science, the Niigata University Global
Circus Program for International Student Exchanges, and also the
Major State 973 Program (Grant No. 2007CB815000) as well as the
National Natural Science Foundation of China (Grant Nos. 10775004
and 10975008).

\end{acknowledgments}

%
%


\newpage

\begin{longtable}{ccccccc}
  \caption{Neutron threshold energy $|\lam|$, total average pairing gap
  $\Del_{uv}^{\text{tot.}}$, pair number $\tilde{N}^{\text{tot.}}$ and pair correlation energy
  $E^{\text{tot.}}_{\text{pair}}$ in the $N=86$ isotones are listed in the first four rows.
  The following rows list properties of the individual low-lying quasiparticle states shown in
  Fig.~\ref{fig:qpspectrumN86II}: the peak energy
  $E_{\text{q.p.}}$ and the width $\Gam$ extracted from the pair number density $\tilde{n}_{lj}(E)$,
  the Hartree-Fock single-particle energy $\vep$ corresponding to the quasiparticle state,
  the pair number $\tilde{N}_{nlj}'$, the pair correlation energy $E_{\text{pair},nlj}'$
  and the effective pairing gap $\Del_{uv,nlj}'$ evaluated for the quasiparticle state
  $lj$ within the energy interval $E=0-4$ MeV.
  The unit of the energy is MeV, except for the width $\Gam$ shown in keV.
  \label{tab:spectrum-N86}}\\
\hline
                          &                            & {$^{136}$Sn}   & {$^{134}$Cd}   & {$^{132}$Pd}   & {$^{130}$Ru}    & {$^{128}$Mo} \\ \hline  \endfirsthead
  \caption{(continued)} \\ \hline
                          &                            & {$^{136}$Sn}   & {$^{134}$Cd}   & {$^{132}$Pd}   & {$^{130}$Ru}    & {$^{128}$Mo} \\ \hline  \endhead
 \multicolumn{2}{c}{$|\lam|$}                          & {{$2.390$}}    & {{$1.884$}}    & {{$1.383$}}    & {{$0.894$}}     & $0.421$\\
 \multicolumn{2}{c}{$\Del_{uv}^{\text{tot.}}$}         & $0.736 $       & $0.721 $       & $0.707 $       & $0.694 $        & $0.678 $  \\
\multicolumn{2}{c}{$\tilde{N}^{\text{tot.}} $} & $16.875$       & $17.083$       & $17.458$       & $18.111$        & $19.245$  \\
 \multicolumn{2}{c}{$E^{\text{tot.}}_{\text{pair}}$}   & $-6.212$       & $-6.162$       & $-6.173$       & $-6.280$        & $-6.527$ \\ \hline
                                     & $\vep$~~                   & $-2.302       $ & $-1.799       $ & $-1.297       $ & $-0.801       $ & $-0.309       $  \\
                                     & $E_{\text{q.p.}}$~~        & $0.760        $ & $0.749        $ & $0.740        $ & $0.734        $ & $0.733        $  \\
                                     & $\Gam$~~                   & $<0.1         $ & $0.1          $ & $<0.1         $ & $2.5          $ & $7.3          $  \\
                       {$2f_{7/2}$~~} & ${\Del}_{uv,nlj}'$~~   & $0.755        $ & $0.745        $ & $0.737        $ & $0.731        $ & $0.727        $   \\
                                     & $\tilde{N}_{nlj}'$~~      & $3.951        $ & $3.954        $ & $3.959        $ & $3.966        $ & $3.973        $ \\
                                     & $E_{\text{pair},nlj}'$~~        & $-1.492       $ & $-1.473       $ & $-1.458       $ & $-1.449       $ & $-1.445       $ \\ \hline
                                     & $\vep$~~                   & $-0.480       $ & $-0.235       $ & $-0.010       $ & $0.190        $ & $0.362        $  \\
                                     & $E_{\text{q.p.}}$~~        & $1.965        $ & $1.685        $ & $1.365        $ & $1.081        $ & $0.803        $  \\
                                     & $\Gam$~~                   & $<0.1         $ & $<0.1         $ & $95.0         $ & $238.8        $ & $396.4        $  \\
                      {$3p_{3/2}$~~}  & ${\Del}_{uv,nlj}'$~~   & $0.591        $ & $0.579        $ & $0.568        $ & $0.554        $ & $0.529        $   \\
                                     & $\tilde{N}_{nlj}'$~~      & $0.528        $ & $0.591        $ & $0.679        $ & $0.813        $ & $1.040        $ \\
                                     & $E_{\text{pair},nlj}'$~~        & $-0.156       $ & $-0.171       $ & $-0.193       $ & $-0.225       $ & $-0.275       $ \\ \hline
                                     & $\vep$~~                   & $0.066        $ & $0.257        $ & $0.423        $ & $0.563        $ & $0.678        $  \\
                                     & $E_{\text{q.p.}}$~~        & $2.405        $ & $2.089        $ & $1.760        $ & $1.422        $ & $1.071        $  \\
                                     & $\Gam$~~                   & $122.3        $ & $264.7        $ & $427.4        $ & $612.2        $ & $802.0        $  \\
                      {$3p_{1/2}$~~}  & ${\Del}_{uv,nlj}'$~~   & $0.570        $ & $0.553        $ & $0.540        $ & $0.522        $ & $0.501        $   \\
                                     & $\tilde{N}_{nlj}'$~~      & $0.179        $ & $0.199        $ & $0.226        $ & $0.268        $ & $0.339        $ \\
                                     & $E_{\text{pair},nlj}'$~~        & $-0.051       $ & $-0.055       $ & $-0.061       $ & $-0.070       $ & $-0.085       $ \\ \hline
                                     & $\vep$~~                   & $0.449        $ & $0.862        $ & $1.237        $ & $1.559        $ & $1.816        $  \\
                                     & $E_{\text{q.p.}}$~~        & $2.919        $ & $2.820        $ & $2.694        $ & $2.540        $ & $2.364        $  \\
                                     & $\Gam$~~                   & $21.7         $ & $50.6         $ & $110.1        $ & $223.8        $ & $431.9        $  \\
                      {$2f_{5/2}$~~}  & ${\Del}_{uv,nlj}'$~~   & $0.736        $ & $0.721        $ & $0.708        $ & $0.696        $ & $0.685        $   \\
                                     & $\tilde{N}_{nlj}'$~~      & $0.690        $ & $0.682        $ & $0.684        $ & $0.701        $ & $0.736        $ \\
                                     & $E_{\text{pair},nlj}'$~~        & $-0.254       $ & $-0.246       $ & $-0.242       $ & $-0.244       $ & $-0.252       $ \\ \hline
                                     & $\vep$~~                   & $0.566        $ & $1.215        $ & $1.869        $ & $2.522        $ & $3.163        $  \\
                                     & $E_{\text{q.p.}}$~~        & $3.051        $ & $3.186        $ & $3.334        $ & $3.492        $ & $3.661        $  \\
                                     & $\Gam$~~                   & $0.3          $ & $0.8          $ & $2.2          $ & $7.0          $ & $24.2         $  \\
                      {$1h_{9/2}$~~}  & ${\Del}_{uv,nlj}'$~~   & $0.766        $ & $0.756        $ & $0.751        $ & $0.748        $ & $0.747        $  \\
                                     & $\tilde{N}_{nlj}'$~~      & $1.214        $ & $1.145        $ & $1.079        $ & $1.016        $ & $0.948        $ \\
                                     & $E_{\text{pair},nlj}'$~~        & $-0.465       $ & $-0.433       $ & $-0.405       $ & $-0.380       $ & $-0.354       $ \\ \hline
\end{longtable}

\begin{longtable}{cccccc}
  \caption{Dependence of the ground state pair correlation and the
 quasiparticle properties in $^{128}$Mo on the pairing interaction strength.
  To control the pairing interaction strength, we vary the parameter $\eta$ of
  DDDI in Eq.~({\ref{eq:DDDI-V}}) as $\eta=0.84,~0.71,~0.62,~0.56$. We list here
  the threshold energy $|\lam|$, the total average pairing gap
  $\Del_{uv}$, and the total pair correlation energy $E_{\text{pair}}$, the
  peak energy $E_{\text{q.p.}}$ and the width $\Gam$ of the lowest two
  quasiparticle resonances for $p_{3/2}, p_{1/2}, f_{5/2}$, and $f_{7/2}$.
  The unit of energy is MeV, except for the width $\Gam$ shown in keV.}
  \label{tab:spectrum-128Mo-fp}\\
    \hline
 \multicolumn{2}{c}{$\eta$}                            & $0.84      $ & $0.71     $ & $0.62     $ & $0.56     $   \\ \hline  \endfirsthead
 \multicolumn{2}{c}{$\eta$}                            & $0.84      $ & $0.71     $ & $0.62     $ & $0.56     $   \\ \hline \endhead
 \multicolumn{2}{c}{$|\lam|$}                          & $0.363     $ & $0.421    $ & $0.572    $ & $0.778    $   \\
 \multicolumn{2}{c}{$\Del^{\text{tot.}}_{uv}$}         & $0.414     $ & $0.678    $ & $1.077    $ & $1.541    $   \\
 \multicolumn{2}{c}{$E^{\text{tot.}}_{\text{pair}}$}   & $-2.860    $ & $-6.527   $ & $-13.902  $ & $-24.800  $  \\ \hline
                                 & $\vep$~~            &$-0.330   $ & $-0.309   $ & $-0.292   $ & $-0.283   $   \\
      $2f_{7/2}$~~               & $E_{\text{q.p.}}$~~ &$0.428    $ & $0.733    $ & $1.146    $ & $1.591    $   \\
                                 & $\Gam$~~            &$4.0      $ & $7.3      $ & $13.9     $ & $24.6     $   \\ \cline{2-6}
                                 & $\vep$~~            &$-25.566  $ & $-25.571  $ & $-25.562  $ & $-25.534  $   \\
      $1f_{7/2}$~~               & $E_{\text{q.p.}}$~~ &$25.204   $ & $25.155   $ & $25.013   $ & $24.812   $   \\
                                 & $\Gam$~~            &$0.4      $ & $0.4      $ & $0.1      $ & $1.6      $   \\ \hline
                                 & $\vep$~~            &$0.355    $ & $0.362    $ & $0.369    $ & $0.373    $   \\
      $3p_{3/2}$~~               & $E_{\text{q.p.}}$~~ &$0.670    $ & $0.803    $ & $1.049    $ & $1.346    $ \\
                                 & $\Gam$~~            &$337.7    $ & $396.4    $ & $492.4    $ & $612.6    $ \\ \cline{2-6}
                                 & $\vep$~~            &$-20.175  $ & $-20.151  $ & $-20.114  $ & $-20.078  $  \\
      $2p_{3/2}$~~               & $E_{\text{q.p.}}$~~ &$19.816   $ & $19.742   $ & $19.579   $ & $19.379   $ \\
                                 & $\Gam$~~            &$0.9      $ & $3.1      $ & $9.1      $ & $19.9     $ \\  \hline
                                 & $\vep$~~            &$0.673    $ & $0.678    $ & $0.682    $ & $0.685    $   \\
      $3p_{1/2}$~~               & $E_{\text{q.p.}}$~~ &$0.962    $ & $1.070    $ & $1.290    $ & $1.569    $   \\
                                 & $\Gam$~~            &$727.4    $ & $802.0    $ & $936.2    $ & $1127.7   $   \\ \cline{2-6}
                                 & $\vep$~~            &$-18.553  $ & $-18.528  $ & $-18.494  $ & $-18.463  $   \\
      $2p_{1/2}$~~               & $E_{\text{q.p.}}$~~ &$18.194   $ & $18.119   $ & $17.96    $ & $17.768   $   \\
                                 & $\Gam$~~            &$2.2      $ & $5.6      $ & $13.5     $ & $25.4     $   \\ \hline
                                 & $\vep$~~            &$1.817    $ & $1.816    $ & $1.814    $ & $1.811    $   \\
      $2f_{5/2}$~~               & $E_{\text{q.p.}}$~~ &$2.260    $ & $2.365    $ & $2.594    $ & $2.899    $   \\
                                 & $\Gam$~~            &$392.9    $ & $431.9    $ & $514.4    $ & $660.0    $   \\ \cline{2-6}
                                 & $\vep$~~            &$-20.970  $ & $-21.007  $ & $-21.040  $ & $-21.048  $   \\
      $1f_{5/2}$~~               & $E_{\text{q.p.}}$~~ &$20.609   $ & $20.595   $ & $20.498   $ & $20.338   $   \\
                                 & $\Gam$~~            &$0.2      $ & $0.5      $ & $2.5      $ & $7.4      $   \\ \hline\hline
\end{longtable}


\end{document}